\documentclass[%
 reprint, 
 amsmath,amssymb,
 aps, physrev,
]{revtex4-2}

\usepackage{graphicx}
\usepackage{dcolumn}
\usepackage{bm}

\usepackage{xcolor}
\usepackage{hyperref}

\usepackage[section]{placeins}

\newcommand{\RNum}[1]{\uppercase\expandafter{\romannumeral #1\relax}}

\begin{document}

\preprint{APS/123-QED}

\title{\textbf{Uncovering coupled ionic-polaronic dynamics and interfacial enhancement in Li$_x$FePO$_4$} 
}%

\author{Fengyu Xie}
\affiliation{
 College of Artificial Intelligence and Data Science, Suzhou Institute for Advanced Research, University of Science and Technology of China, Suzhou, Jiangsu 215123, China
}%

\author{Yuxiang Gao}
\affiliation{
 College of Artificial Intelligence and Data Science, Suzhou Institute for Advanced Research, University of Science and Technology of China, Suzhou, Jiangsu 215123, China
}%

\author{Ruoyu Wang}
\affiliation{
 College of Artificial Intelligence and Data Science, Suzhou Institute for Advanced Research, University of Science and Technology of China, Suzhou, Jiangsu 215123, China
}%

\author{Zhicheng Zhong}
\email{zczhong@ustc.edu.cn}
\affiliation{
 College of Artificial Intelligence and Data Science, Suzhou Institute for Advanced Research, University of Science and Technology of China, Suzhou, Jiangsu 215123, China
}
\affiliation{
 Suzhou Lab, Suzhou, Jiangsu 215123, China
}

\date{\today}

\begin{abstract}
Understanding and controlling coupled ionic-polaronic dynamics is crucial for optimizing electrochemical performance in battery materials. However, studying such coupled dynamics remains challenging due to the intricate interplay between Li-ion configurations, polaron charge ordering, and lattice vibrations. Here, we develop a fine-tuned machine-learned force field (MLFF) for Li$_x$FePO$_4$ that captures coupled ion-polaron behavior. Our simulations reveal picosecond-scale polaron flips occurring orders of magnitude faster than Li-ion migration, featuring strong correlation to Li configurations. Notably, polaron charge fluctuations are further enhanced at Li-rich/Li-poor phase boundaries, suggesting a potential interfacial electronic conduction mechanism. These results demonstrate the capability of fine-tuned MLFFs to resolve complex coupled transport, and provide insight into emergent ionic-polaronic dynamics in multivalent battery cathodes.
\end{abstract}

\maketitle


\textit{Introduction.}
Battery materials have become a central focus in materials science, and understanding their microscopic working mechanisms is key to improving electrochemical performance. In typical electrode materials such as Li$_x$FePO$_4$, partial delithiation ($0<x<1$) gives rise to both Li–vacancy ordering and multiple valence states of transition metals, manifested as small polarons, i.e., localized charge carriers coupled with lattice distortion~\cite{frohlichElectronsLatticeFields1954, StudiesPolaronMotion1959, reticcioliSmallPolaronsTransition2020}. In Li$_x$FePO$_4$, electron and hole polarons localize as Fe$^{2+}$/Fe$^{3+}$ species accompanied by local expansion and contraction of the FeO$_6$ octahedra~\cite{zaghibElectronicOpticalMagnetic2007, ellisSmallPolaronHopping2006}. Charge transport in such materials involves thermally activated hopping: Li ions migrate between vacant and occupied sites, while electronic conductivity arises from polaron hopping between neighboring Fe sites, mediated by lattice vibrations, rather than band-like transport as in metals or doped semiconductors. The two processes are strongly coupled through the interplay of ionic configuration, charge ordering, and lattice distortion, resulting in high structural and dynamical complexity. Despite extensive interest, a clear picture of such coupled ionic–polaronic dynamics remains elusive, even in a model system like Li$_x$FePO$_4$. Fundamental questions, such as the relative contributions of ionic and electronic conductivities~\cite{wangIonicElectronicConducting2007, aminAnisotropyElectronicIonic2007} and the nature of its phase separation during cycling~\cite{delmasLithiumDeintercalationLiFePO42008, delacourtExistenceTemperaturedrivenSolid2005, malikKineticsNonequilibriumLithium2011}, are still under debate. Progress has been hindered by intrinsic complexity of interactions and extrinsic effects such as grain boundaries and defects.

Ab-initio molecular dynamics (AIMD) in principle allows direct access to coupled ionic-polaronic transport, but its high computational cost restricts first-principles studies to small supercells and static calculations, such as phase diagrams~\cite{zhouConfigurationalElectronicEntropy2006} or migration barriers~\cite{morganLiConductivityLisub2004, maxischInitioStudyMigration2006, ongComparisonSmallPolaron2011, nakayamaDensityFunctionalStudies2016, wangDecisiveRoleElectrostatic2023}. While recent machine-learned force fields (MLFFs) have enabled large-scale atomistic simulations~\cite{NEURIPS2022_4a36c3c5, rhodesOrbv3AtomisticSimulation2025, chenUniversalGraphDeep2022a, zhangDPA2LargeAtomic2024}, general-purpose models often fail to resolve the subtle structural signatures associated with polaron formation. Modeling the energetics of polarons and their coupling with complex Li–vacancy ordering remains challenging, yet it is essential for capturing coupled ionic–polaronic dynamics. Specialized models like CHGNet~\cite{dengCHGNetPretrainedUniversal2023} and Leopold~\cite{birschitzkyMachineLearningSmall2025} incorporate explicit spin or charge labels to track valence states, but typically require customized architectures, limiting compatibility with existing pretrained frameworks.

In our recent work~\cite{wangPFDAutomaticallyGenerating2025}, we showed that by fine-tuning a pretrained general-purpose MLFF on a system-specific dataset, it is possible to achieve meV/atom-level accuracy sufficient to capture polaronic behavior. Building on this, we develop a near-DFT-accurate MLFF for Li$_x$FePO$_4$ that distinguishes Fe valence states purely through local structural features such as Fe–O bond lengths, without explicit charge labeling. Our model captures key dynamical features of polarons, including picosecond-scale polaron flips and their strong correlation with Li-vacancy configurations. Notably, we observe a pronounced enhancement of polaron flipping near Li-rich/Li-poor phase boundaries, suggesting a potential interfacial electronic conduction mechanism. These results open new opportunities for efficiently modeling coupled ionic–polaronic dynamics in complex cathode materials, and provide physical insights into their emergent charge transport mechanisms.

\begin{figure*}[htbp]
    \centering
    \includegraphics[width=1.0\linewidth]{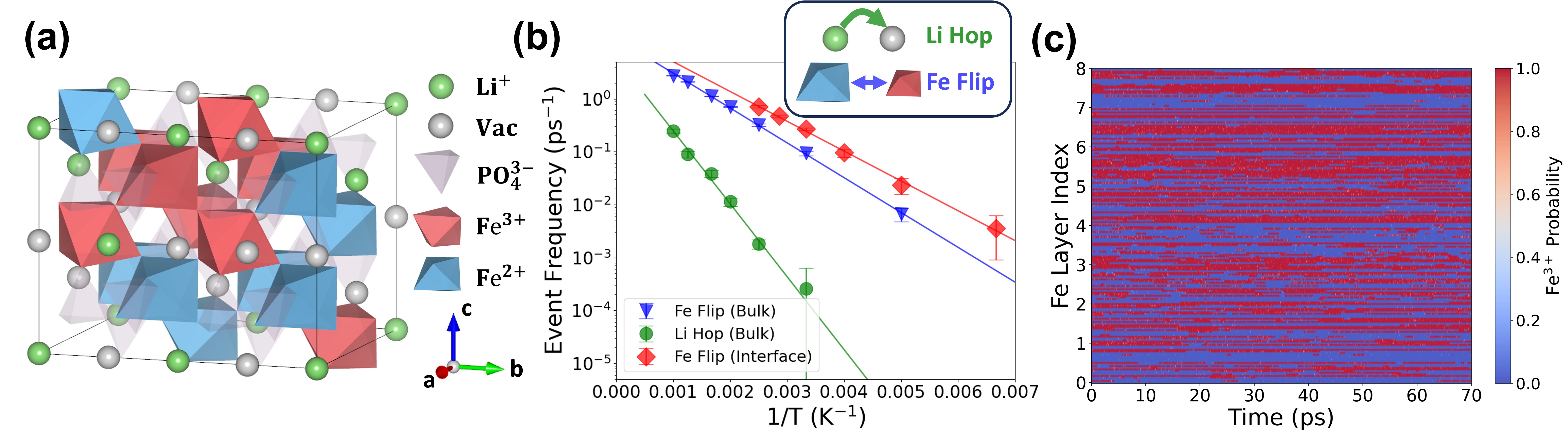}
    \caption{(a) A $1\times2\times2$ supercell structure of olivine Li$_x$FePO$_4$. Green spheres indicate Li$^+$ ions. Purple tetrahedra represent PO$_4^{3-}$ ions. Blue and red octahedra represent Fe$^{2+}$O$_6$ and Fe$^{3+}$O$_6$ coordination environments, respectively. (b) Arrhenius plot of bulk Fe polaron flip rate (blue), bulk Li hop rate (green), and interfacial Fe polaron flip rate (red) as a function of inverse temperature ($1/T$). The bulk values were computed in a disordered solid-solution. The event frequencies are plotted on a logarithmic scale along the y-axis. (c) Time evolution of Fe valence states projected onto each Fe layer for a random Li ordering at $x=0.5$, $T=300$K. Color indicates the probability of Fe$^{3+}$ character as determined by a Gaussian mixture classifier; red denotes a higher Fe$^{3+}$ probability.}
    \label{fig:fig1}
\end{figure*}

\textit{Methods.}
In this work, we study Li$_x$FePO$_4$ in its orthorhombic olivine phase~\cite{MaterialsDataLiFePO42020} ($a=10.236~\mathrm{\AA}$, $b=5.971~\mathrm{\AA}$, $c=4.655~\mathrm{\AA}$, corresponding to the $x$, $y$ and $z$ directions, respectively), which contains 4 formula units per primitive cell. Figure~\ref{fig:fig1}(a) illustrates a $1\times2\times2$ supercell. Li ions occupy 4a octahedral sites and migrate along edge-sharing channels in the $y$-direction, while Fe$^{2+}$/Fe$^{3+}$ ions reside on corner-sharing 4c sites. Polaron flips between these valence states manifest as local expansion or contraction of FeO$_6$ octahedra, with Fe$^{3+}$ exhibiting shorter Fe–O bonds due to stronger electrostatic attraction. Although our MLFF lacks explicit charge tracking, Fe–O bond lengths allow indirect identification of polaron flips, which serve as a proxy for polaron dynamics and its coupling to lattice fluctuations.

A pretrained DPA-2 MLFF~\cite{zhangDPA2LargeAtomic2024} was fine-tuned on a DFT dataset across various Li contents ($x$) and representative configurations of Li, vacancies, and polarons. The resulting model achieves a root-mean-square error of $\sim$2.1 meV/atom in energy and $\sim58.5~\mathrm{meV/\AA}$ in force. Readers are referred to Supplementary Section \RNum{1} for additional details. All DFT data\cite{Datasets351LixFePO4_trajectory_plus_aimd} and the fine-tuned MLFF\cite{Models350DPA231LFPOfinetuned} are available on AIS Square.

Using Fe–O bond lengths as a structural fingerprint of oxidation states, we show that the fine-tuned model accurately reproduces the geometry and energetics of polaronic configurations (Supplementary Section \RNum{2} A). In MD simulations, Fe$^{2+}$/Fe$^{3+}$ states persist dynamically without thermal collapse across a wide range of temperatures and compositions (Supplementary Section \RNum{2} B). This behavior is consistent with previous proposals that valence ordering contributes an additional configurational entropy term, which helps stabilize the solid-solution phase at elevated temperatures~\cite{delacourtExistenceTemperaturedrivenSolid2005, zhouConfigurationalElectronicEntropy2006}, thereby showing our method as a reliable framework for investigating polaron dynamics.

\textit{Results.} 
We now present the results obtained using our fine-tuned MLFF. First, we show that polaron flips occur on picosecond timescales, orders of magnitude faster than Li-ion migration. We then analyze the spatial and temporal distribution of polaron flip events, finding that they exhibit short-ranged, transient, and charge-compensating correlations, and their frequency strongly couple with Li–vacancy ordering. Finally, we examine a LiFePO$_4$/FePO$_4$ interface and find that polaronic activity is notably enhanced near the phase boundary.

Figure~\ref{fig:fig1}(b) compares the frequency of Fe polaron flip (per Fe atom) with that of Li hopping events (per Li atom) over the temperature range of 300–1000 K in a disordered solid solution at $x=0.5$. Fe$^{2+}$ and Fe$^{3+}$ coexist in a dynamic equilibrium, with polaron fluctuations occurring orders of magnitude faster than Li-ion migration, approaching a picosecond regime (Figure~\ref{fig:fig1}(b)-(c)). The effective activation energy for polaron flipping is $0.131 \pm 0.002$ eV, notably lower than Li migration barrier, $0.273 \pm 0.016$ eV (Figure~\ref{fig:fig1}(b) and Table~\ref{tab:tab1}). This finding aligns with single-crystal measurements~\cite{aminAnisotropyElectronicIonic2007} and theoretical studies~\cite{nakayamaDensityFunctionalStudies2016}, which indicate that in crystalline Li$_x$FePO$_4$, rate performance is limited by Li-ion mobility rather than electronic conductivity, contrary to conclusions drawn from polycrystalline samples\cite{wangIonicElectronicConducting2007}.

\begin{table}[htbp]
\caption{\label{tab:tab1}%
Prefactor and effective activation energy in Arrhenius equation by fitting the rate of bulk Fe polaron flips, bulk Li hops and interfacial Fe polaron flips (Figure~\ref{fig:fig1}) with $\mathrm{ln}(r)=\mathrm{ln}(A)-E_a/k_{\rm B}T$.
}
\begin{ruledtabular}
\begin{tabular}{lcc}
\textrm{Event type}&
\textrm{$A$ (ps$^{-1}$)}&
\textrm{$E_a$ (eV)}\\
\colrule
Li hop (bulk)         & 5.9  & $0.273\pm0.016$\\
Fe flip (bulk)        & 13.9 & $0.131\pm0.002$\\
Fe flip (interfacial) & 18.4 & $0.112\pm0.003$\\
\end{tabular}
\end{ruledtabular}
\end{table}

\begin{figure}[tbp]
    \centering
    \includegraphics[width=1.0\linewidth]{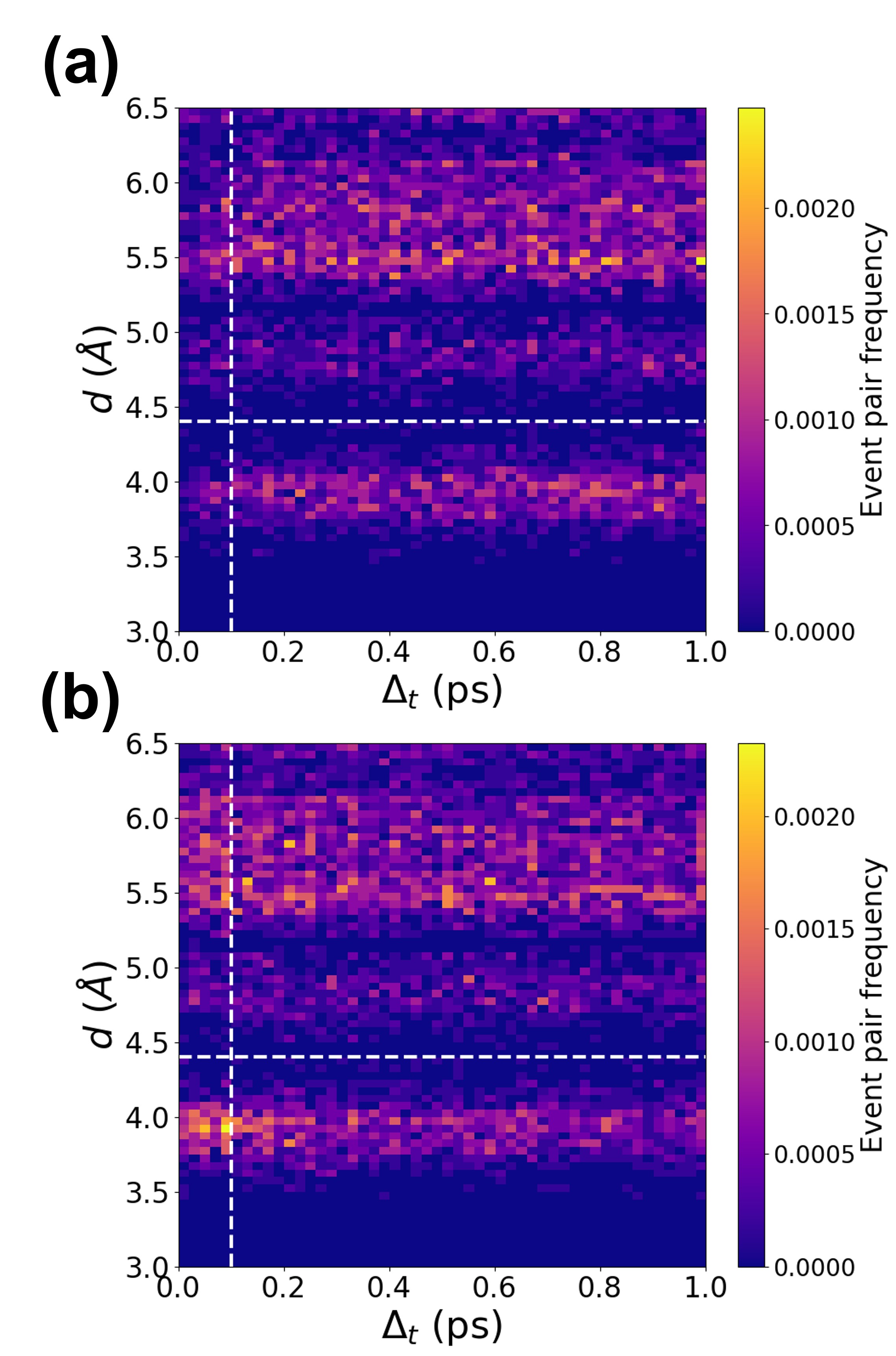}
    \caption{Densities of polaron flip event pairs extracted from MD simulations at $x = 0.5$ and $T = 400$ K in a disordered bulk configuration. Frequencies of polaron flip event pairs are binned according to their spatial distance ($d = 3.0\sim6.5~\mathrm{\AA}$) and time lag ($\Delta t = 0\sim1.0$ 
    ps) between the two events. (a) Same-sign pairs ([Fe$^{2+} \to$ Fe$^{3+}$, Fe$^{2+} \to$ Fe$^{3+}$] and [Fe$^{3+} \to$ Fe$^{2+}$, Fe$^{3+} \to$ Fe$^{2+}$]); (b) Opposite-sign pairs ([Fe$^{2+} \to$ Fe$^{3+}$, Fe$^{3+} \to$ Fe$^{2+}$]). Brighter colors indicate higher densities of event pairs occurring at a given spatial and temporal separation. The vertical and horizontal dashed lines mark $\Delta t=0.1$ ps and $d=4.4~\mathrm{\AA}$, respectively.}
    \label{fig:fig2}
\end{figure}

We further demonstrate how Fe polaron flip events are spatially and temporally coupled. Without explicit charge labels, individual polaron trajectories cannot be resolved. Still, polaron flip statistics (Table~\ref{tab:tab2} and Figure~\ref{fig:fig2}) reveal short-time, short-range charge compensation behavior, with opposite-sign pairs of flips ([Fe$^{2+}\rightarrow$Fe$^{3+}$, Fe$^{3+}\rightarrow$Fe$^{2+}$]) enhanced at $\Delta t \leq 0.05$ ps and Fe–Fe distances below $4.4~\mathrm{\AA}$ (corresponding to the nearest-neighbor distance between Fe atoms in the $yz$ plane) compared to same-sign pairs ([Fe$^{2+}\rightarrow$Fe$^{3+}$, Fe$^{2+}\rightarrow$Fe$^{3+}$] or [Fe$^{3+}\rightarrow$Fe$^{2+}$, Fe$^{3+}\rightarrow$Fe$^{2+}$]), suggesting short-range, transient polaron migration.

\begin{table}[htbp]
\caption{\label{tab:tab2}%
Preference ratio ($\mathrm{PR}$), defined as the ratio of opposite-sign to same-sign polaron flip event frequencies, measured across different spatial ($d$) and temporal ($\Delta t$) distances.
Data are extracted from the same MD simulation as in Figure~\ref{fig:fig2}.
}
\begin{ruledtabular}
\begin{tabular}{l|cc}
\textrm{$\mathrm{PR}=r_{\mathrm{opp}}/r_{\mathrm{same}}$}&
\textrm{$d\leq4.4~\mathrm{\AA}$}&
\textrm{$d > 4.4~\mathrm{\AA}$}\\
\colrule
$\Delta t \leq 0.05$ ps  & 3.971  & 2.212\\
$\Delta t > 0.05$ ps     & 1.000 & 0.999\\
\end{tabular}
\end{ruledtabular}
\end{table}

\begin{figure}[htbp]
    \centering
    \includegraphics[width=1.0\linewidth]{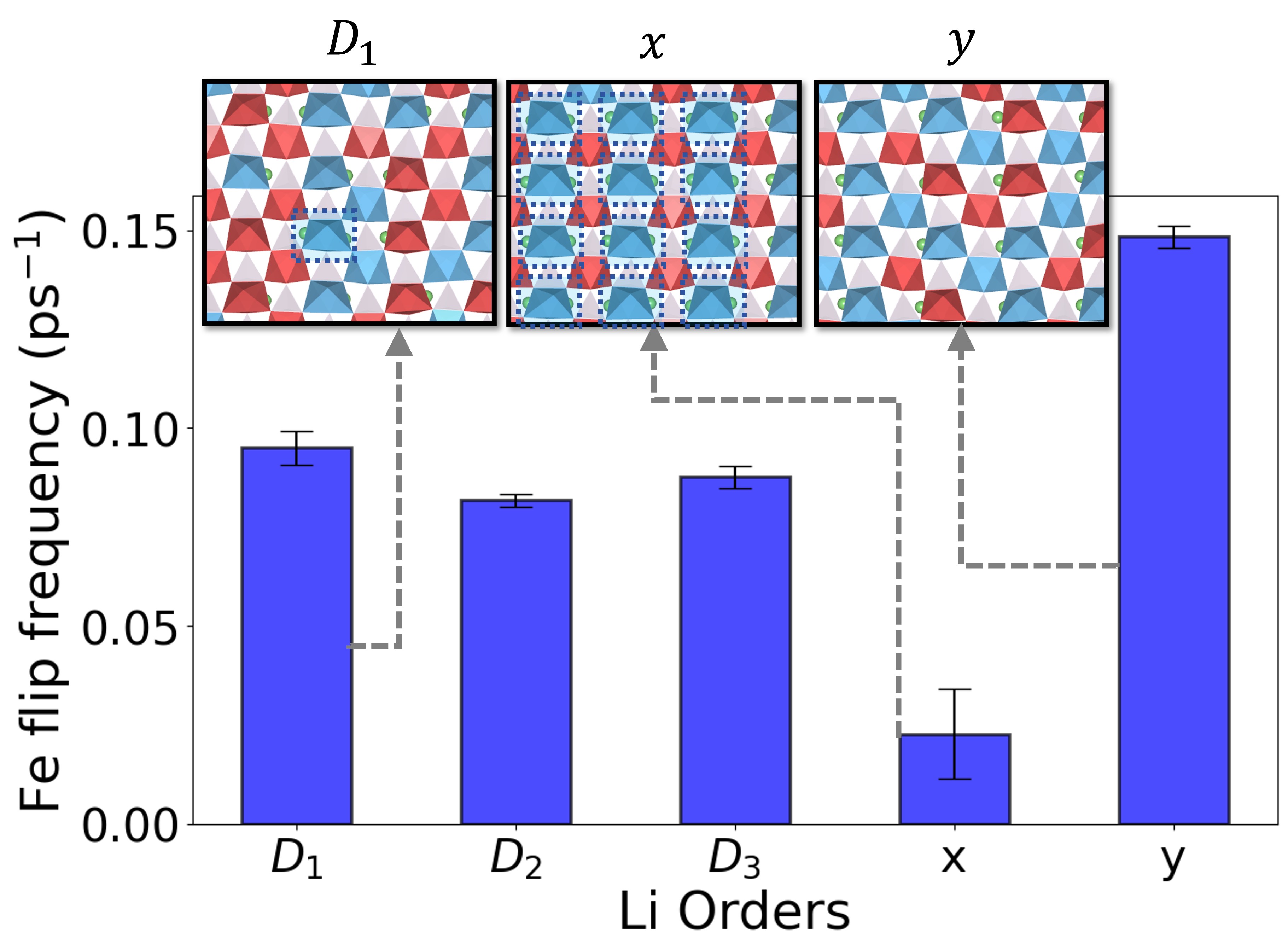}
    \caption{Computed Fe polaron flip frequency (per Fe atom) for 5 different Li ordering configurations in solid solution. The panels above bars display snapshots of polaron charge orders on the $yz$ plane from MD simulations in the $D_1$(left), $x$(middle) and $y$(right) configurations. Fe$^{2+}$ and Fe$^{3+}$ ions are represented as blue and red octahedra, respectively. Blue boxes with dashed borders indicate the Li-Fe-Li clamp motif along the $y$ direction.}
    \label{fig:fig3}
\end{figure}

\begin{figure*}[htbp]
    \centering
    \includegraphics[width=1.0\linewidth]{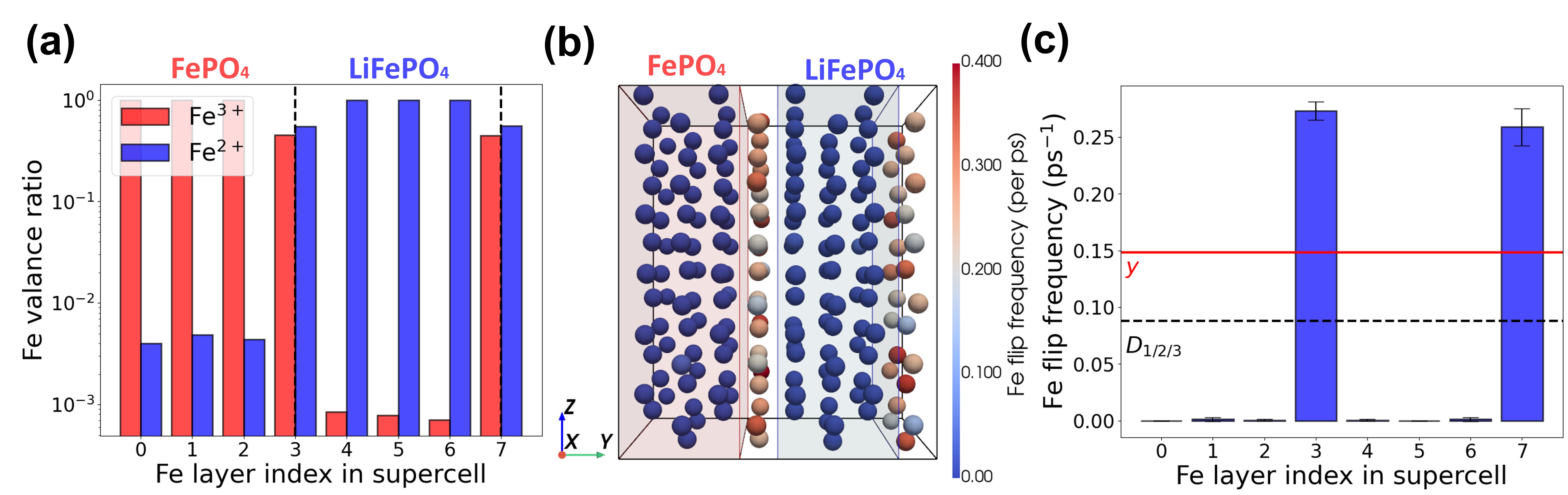}
    \caption{(a) Layer-resolved Fe valence ratio along the $y$-direction, plotted on a logarithmic scale to highlight valence mixing across the phase boundary. Red and blue bars represent Fe$^{3+}$ and Fe$^{2+}$, respectively. (b) Schematic of polaron flip frequencies on each Fe site across the supercell. The FePO$_4$ and LiFePO$_4$ domains are shaded in red and blue, respectively. Red spheres indicate higher flip frequency on site. (c) Layer-by-layer Fe flip frequency (per Fe ion). The red line marks the bulk value in a $y$-ordered solid solution; the dashed black line denotes the average flip frequency over three disordered bulk configurations ($D_1$, $D_2$, $D_3$).}
    \label{fig:fig4}
\end{figure*}

To examine how Li-vacancy configurations affect polaron dynamics, we performed MD simulations at $x=0.5$ and $T = 300$ K using supercells with 5 distinct Li orderings. These include three chemically disordered configurations ($D_1$, $D_2$, and $D_3$), as well as two ordered configurations in which half of the Li ions are removed from alternating planes: one along the $x$-direction ($x$-order, alternating $yz$ planes) and the other along the $y$-direction ($y$-order, alternating $xz$ planes). Figure~\ref{fig:fig3} compares Fe polaron flip rates under these Li-vacancy orderings. The $x$-order (middle panel) strongly suppresses flipping due to a local Li–Fe–Li clamp motif (highlighted in blue dashed boxes) formed along the $y$ direction, which stabilizes electron polaron (i.e., Fe$^{2+}$) via electrostatic attraction and penalizes polaron disorder. In contrast, the $y$-order (right panel) fully removes such constraints, resulting in a flatter energy landscape and enhanced polaron activity. Disordered configurations (left panel) partially disrupt the clamp motif, yielding intermediate flipping rates. We notice that these artificial orderings may not correspond to thermodynamic equilibrium states in bulk systems at room temperature, where phase separation is generally expected~\cite{yamadaPhaseChangeLisub2005, doddPhaseDiagramLixFePO42006}. Nonetheless, previous works~\cite{malikKineticsNonequilibriumLithium2011, liuCapturingMetastableStructures2014, orikasaDirectObservationMetastable2013, xiaoKineticMonteCarlo2018} have shown that under certain conditions, such as in nanoscale particles or during fast dynamic charge/discharge cycles, LiFePO$_4$ may transiently form metastable solid-solution. In that context, our results suggest that Li-vacancy ordering can play a critical role in polaron dynamics, and this interplay between ion and polaron transport warrants attention.

At room temperature, Li$_x$FePO$_4$ is known to undergo phase separation into Li-rich and Li-poor domains, giving rise to interfaces between coexisting LiFePO$_4$ and FePO$_4$~\cite{yamadaPhaseChangeLisub2005, doddPhaseDiagramLixFePO42006}. To explore the behavior of small polarons near such interfaces, we constructed a supercell comprising adjacent LiFePO$_4$ ($x = 1.0$) and FePO$_4$ ($x = 0.0$) regions, separated along the $y$-axis, and performed MD simulations at 300 K. Figure~\ref{fig:fig4}(a) shows the distribution of Fe valence ratios projected along the $y$-direction. As expected, Fe$^{2+}$ (blue) and Fe$^{3+}$ (red) are nearly entirely confined to the Li-rich and Li-poor regions, respectively, with only less than 0.5\% intermixing. However, within the interfacial layers, we observe nearly equal proportions of Fe$^{2+}$ and Fe$^{3+}$, suggesting  dynamic coexistence of valence states at the interface. As illustrated in Figure~\ref{fig:fig4}(b), unlike in a disordered solid solution, valence-flipping events are predominantly localized at the interface. In the bulk regions of LiFePO$_4$ and FePO$_4$, the absence of adjacent Fe ions with different valences likely restricts polaron migration. In contrast, the interface contains both Fe$^{2+}$ and Fe$^{3+}$, enabling polaron exchange.

Remarkably, we find that polaron flips occur at even higher rates at the interface than in the fastest solid solution configuration (i.e., the $y$-order, red solid line), as shown in Figure~\ref{fig:fig4}(c). While this behavior may not fully represent the steady-state dynamics in real, phase-separated Li$_x$FePO$_4$, particularly considering that real interfaces can be rough and defective, it nevertheless highlights a physically intriguing scenario in which phase boundaries act as high-conductivity channels for small polaron transport. Similar enhanced carrier dynamics have been discovered in other oxide interfaces, such as LaAlO$_3$/SrTiO$_3$~\cite{kongFormationTwodimensionalSmall2019, thielTunableQuasiTwoDimensionalElectron2006, mannhartOxideInterfacesOpportunity2010, ohtomoHighmobilityElectronGas2004}.

The observed interfacial enhancement may be attributed to several physical factors. One likely cause is the lattice mismatch along the $y$-direction between the Li-rich and Li-poor domains, which induces local interfacial strain. This strain could soften specific vibrational modes localized near the interface, thereby lowering the effective energy barrier for polaron migration. As shown in Figure~\ref{fig:fig1}(b) and Table~\ref{tab:tab1}, the interface exhibits a lower effective activation energy ($E_a = 0.112 \pm 0.003$ eV) for polaron flipping compared to the disordered bulk ($E_a = 0.131 \pm 0.002$ eV), suggesting the presence of a more favorable dynamic environment. Furthermore, the geometry of the interface naturally reduces the effective dimensionality. Since the phonon density of states $g(\omega)$ in the acoustic regime increases with reduced dimensionality and scales as $\omega^{d-1}$, the enhanced availability of low-energy modes may facilitate thermally activated flipping if such modes effectively couple with local valence fluctuations. In accordance with this hypothesis, we find the fitted Arrhenius kinetic pre-factor $A$ (Table~\ref{tab:tab1}) of polaron flip events among interfacial Fe ions to be 18.4 ps$^{-1}$, higher than in disordered bulk (13.9 ps$^{-1}$). 

Despite the insights gained, our approach inherits limitations from the DFT framework and the Born–Oppenheimer approximation, modeling valence dynamics solely on ground-state potential energy surfaces and neglecting non-adiabatic effects. This may contribute to the underestimation of polaron activation barriers relative to experiments~\cite{aminAnisotropyElectronicIonic2007, delacourtUnderstandingElectricalLimitations2005,xuElectronicStructureElectrical2004}. Structurally, the lack of corner-sharing FeO$_6$ octahedra within $yz$-planes disfavors in-plane polaron hopping. Flips at interfaces may involve transient detours and returns along $y$-direction Fe–O–Fe chains, though such pathways remain speculative.

\textit{Conclusion.}
In summary, we investigated coupled ionic-polaronic dynamics in the multivalent transition-metal oxide Li$_x$FePO$_4$ using a fine-tuned machine-learned force field. The model reproduces polaronic geometries and energetics with near-DFT accuracy, enabling valence-state identification through Fe–O bond lengths across large-scale molecular dynamics trajectories. We show that Fe polaron flips occur on picosecond timescales, orders of magnitude faster than Li-ion migration, indicating that charge transport in the single-crystal solid-solution regime is limited by ionic rather than electronic mobility. Polaron activity is strongly modulated by Li–vacancy ordering, and is significantly enhanced at the LiFePO$_4$/FePO$_4$ phase boundary, pointing to a possible interfacial electronic conduction channel. While constrained by DFT limitations and the absence of explicit charge labels, our model captures key thermodynamic and kinetic features of polaron dynamics, demonstrates the power of fine-tuned MLFFs for exploring coupled ion–polaron mechanisms in complex battery materials, and provides physical insights into their charge transport.

\textit{Acknowledgments.}
This work was supported by the National Key R\&D Program of China (Grants No. 2021YFA0718900) and National Nature Science Foundation of China (Grants No. 12374096 and No. 92477114). We thank DP Technology for providing computational resources through the Bohrium platform and data hosting services via AIS Square.


\bibliography{refs}

\end{document}


\preprint{APS/123-QED}

\title{\textbf{Supplementary Information: Uncovering coupled ionic-polaronic dynamics and interfacial enhancement in Li$_x$FePO$_4$} 
}%

\author{Fengyu Xie}
\affiliation{
 College of Artificial Intelligence and Data Science, Suzhou Institute for Advanced Research, University of Science and Technology of China, Suzhou, Jiangsu 215123, China
}%

\author{Yuxiang Gao}
\affiliation{
 College of Artificial Intelligence and Data Science, Suzhou Institute for Advanced Research, University of Science and Technology of China, Suzhou, Jiangsu 215123, China
}%

\author{Ruoyu Wang}
\affiliation{
 College of Artificial Intelligence and Data Science, Suzhou Institute for Advanced Research, University of Science and Technology of China, Suzhou, Jiangsu 215123, China
}%

\author{Zhicheng Zhong}
\email{zczhong@ustc.edu.cn}
\affiliation{
 College of Artificial Intelligence and Data Science, Suzhou Institute for Advanced Research, University of Science and Technology of China, Suzhou, Jiangsu 215123, China
}
\affiliation{
 Suzhou Lab, Suzhou, Jiangsu 215123, China
}

\date{\today}

\maketitle

\section{\label{sec:computational}Computational Details}
\subsection{\label{sec:dft}DFT Computations}
All DFT calculations, including structural relaxations and AIMD simulations, were performed using the Vienna Ab initio Simulation Package (VASP) with the projector augmented-wave (PAW) method~\cite{kresseInitioMolecularDynamics1993, kresseEfficiencyAbinitioTotal1996, kresseEfficientIterativeSchemes1996, kresseUltrasoftPseudopotentialsProjector1999}. A plane-wave energy cutoff of 520 eV and a $\Gamma$-centered $k$-point mesh with spacing of 0.4 $\mathrm{\AA}^{-1}$ were used. Li 1s and Fe 3p electrons were treated as valence electrons. Electronic and ionic relaxations were converged to $10^{-6}$ eV and 0.02 eV/$\mathrm{\AA}$, respectively. The Perdew-Burke-Ernzerhof (PBE) functional~\cite{perdewGeneralizedGradientApproximation1996, perdewGeneralizedGradientApproximation1997} was used with a Hubbard $U$ correction of 5.3 eV on Fe-$3d$ orbitals~\cite{dudarevElectronenergylossSpectraStructural1998a, jainCommentaryMaterialsProject2013, HubbardValuesMaterials2023a}, assuming collinear ferromagnetic order.

Two types of DFT-labeled frames were prepared:
\begin{itemize}
  \item \textbf{Relaxation trajectories:} 126 distinct $1\times2\times2$ Li$_x$FePO$_4$ supercells (16 formula units, extended from an orthorhombic unit cell with $a=10.23620~\mathrm{\AA}$,  $b=5.97076~\mathrm{\AA}$,  $c=4.65492~\mathrm{\AA}$) were initialized with randomized Li, vacancy, and polaron configurations at various $x$ values, then relaxed. A total of 9379 electronically converged frames were generated. From these, 20 configurations were selected, each contributing 41 randomly sampled frames for training.
  
  \item \textbf{AIMD trajectories:} Constant-temperature AIMD simulations at 550 K were conducted on nine $1\times2\times2$ supercells at $x =$ 0.0 to 1.0. Each trajectory lasted 1 ps with a timestep of 4 fs, yielding 1462 converged frames. Five compositions ($x =$ 0.0, 0.25, $0.50$, 0.75, 1.00) were selected, with 150 frames randomly sampled from each.
\end{itemize}

In total, 1570 frames covering diverse Fe valence and Li-vacancy configurations were used to train the MLFF.

\subsection{\label{sec:ft}MLFF Fine-tuning}

The DPA-2 architecture~\cite{zhangDPA2LargeAtomic2024} was adopted and fine-tuned using the DeepMD-kit v3 package~\cite{zengDeePMDkitV3MultipleBackend2025}. The base model (DPA-2.3.1-v3.0.0rc0) was obtained from AIS Square~\cite{Models287DPA231v300rc0}, and fine-tuned on the \textit{Domains\_Anode} model branch. Only energy and force labels were used; no charge, oxidation state, or spin information was included.

\begin{figure}[htbp]
    \centering
    \includegraphics[width=0.9\linewidth]{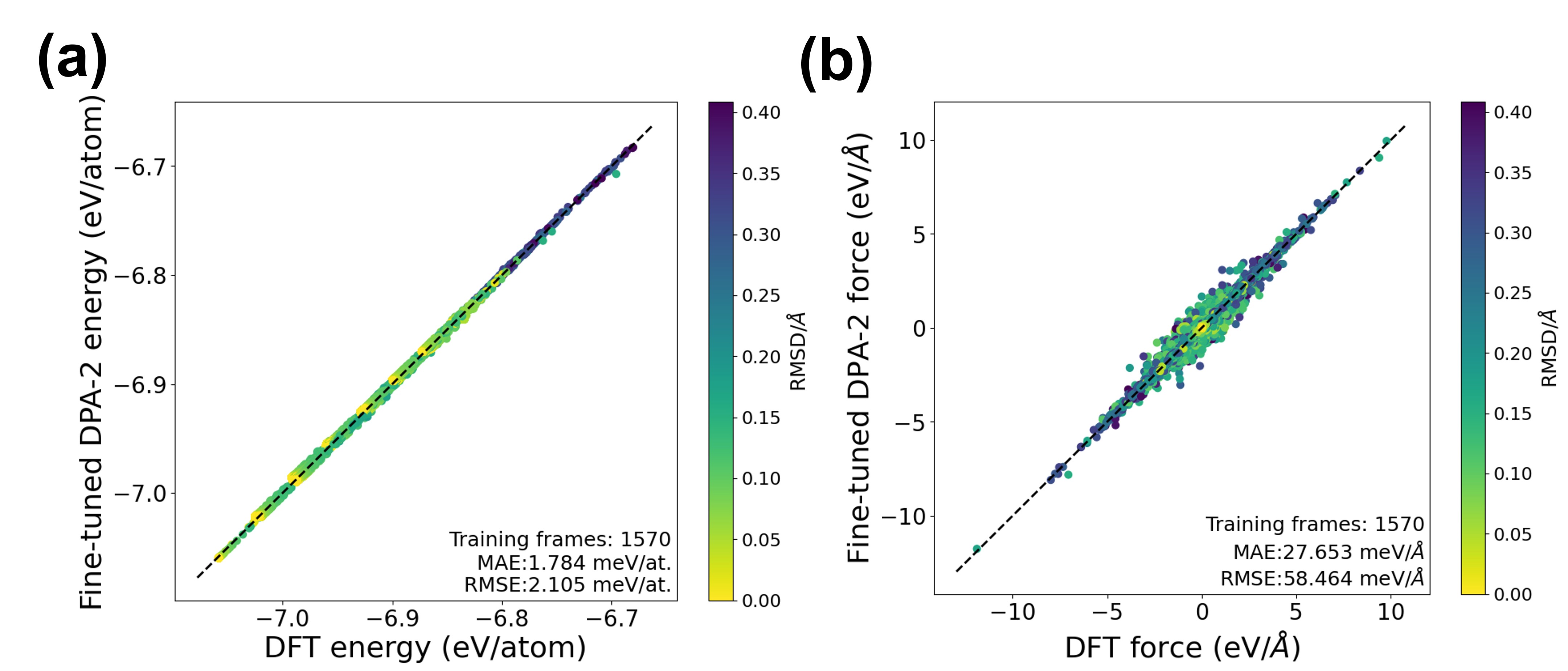}
    \caption{Parity plots showing the errors in (a) energy per atom and (b) atomic forces during the fine-tuning of DPA-2 using DFT reference data. The plots include all 10841 structural frames described in Section~\ref{sec:dft}. The color bar indicates the root-mean-squared deviation (RMSD) of each structure from its corresponding DFT-relaxed minimum; darker colors represent structures with larger deviations from equilibrium geometry.}
    \label{fig:fig-s1}
\end{figure}

Training was performed for 120 epochs. The energy loss weight increased from 0.02 to 1.0, and the force weight decreased from 1000 to 1.0. The learning rate started at 0.001 and decayed by a factor of 0.526 every 11750 steps, reaching $3.51 \times 10^{-8}$. The final model achieved an RMSE of $\sim$2.1 meV/atom in energy and $\sim$58.5 meV/$\mathrm{\AA}$ in force over the full dataset (Figure~\ref{fig:fig-s1}).

\subsection{\label{sec:md}MD Simulations}

MD simulations were performed using the fine-tuned model in LAMMPS~\cite{thompsonLAMMPSFlexibleSimulation2022}, with the NPT ensemble at $P=1$ bar. Temperature and pressure were controlled using damping parameters of 0.1 ps$^{-1}$ and 1.0 ps$^{-1}$, respectively, with a timestep of 1 fs. All simulations used $2\times4\times6$ supercells (192 formula units).

For temperature-dependent analysis, simulations were performed at $x=$ 0.25, 0.50, and 0.75. Each was cooled from 1000 K to 20 K in steps ($T =$ 1000, 800, 600, 500, 400, 300, 200, 100, 20 K), followed by heating back to 1000 K and a second cooling cycle. Simulation durations per temperature ranged from 10 to 30 ps, with the first 1/3 discarded as non-equilibrated. Averages were taken across cycles.

For Li ordering analysis, additional 80 ps simulations at $x=0.5$ and $T=300$ K were performed for the $D_1$, $D_2$, $D_3$, $x$ and $z$ configurations. The first 1/8 of each trajectory was excluded from analysis. The Li orders were preserved throughout the simulations as no Li migration happened at 300 K from visually inspecting the trajectory.

\subsection{\label{sec:post}Post-processing}

Fe valence states, flip frequencies, and hop frequencies were extracted from MD trajectories. The valence of each Fe atom was inferred from its average Fe–O bond length, using a Gaussian mixture model (GMM) fitted at each composition and temperature. A decision boundary was drawn at equal-probability, and classification accuracy was quantified by the Bayes error rate (BE)~\cite{dudaPatternClassification2001}:
\begin{equation}\label{eq:bayes}
    \mathrm{BE} = \mathbb{E}_{x}\!\bigl[\,1-P(\hat C(x)\mid x)\bigr],
\end{equation}
where $\hat{C}(x)$ is the true label and $P(\hat{C}(x)|x)$ the assigned probability. Lower $BE$ indicates better distinguishability between Fe$^{2+}$ and Fe$^{3+}$.

Li hops were identified as jumps between neighboring octahedral sites along $y$ within 0.01 ps. Valence flip events were defined as Fe$^{2+} \leftrightarrow$ Fe$^{3+}$ transitions between adjacent frames, smoothed with a temporal Gaussian filter to suppress thermal noise. A minimum lifetime of 0.08 ps was required for a valence state to be recognized as a valid transition.

\section{Supplementary Results}
\subsection{\label{sec:validation}Validation of polaron configuration energetics}
\begin{figure}[htbp]
    \centering
    \includegraphics[width=0.85\linewidth]{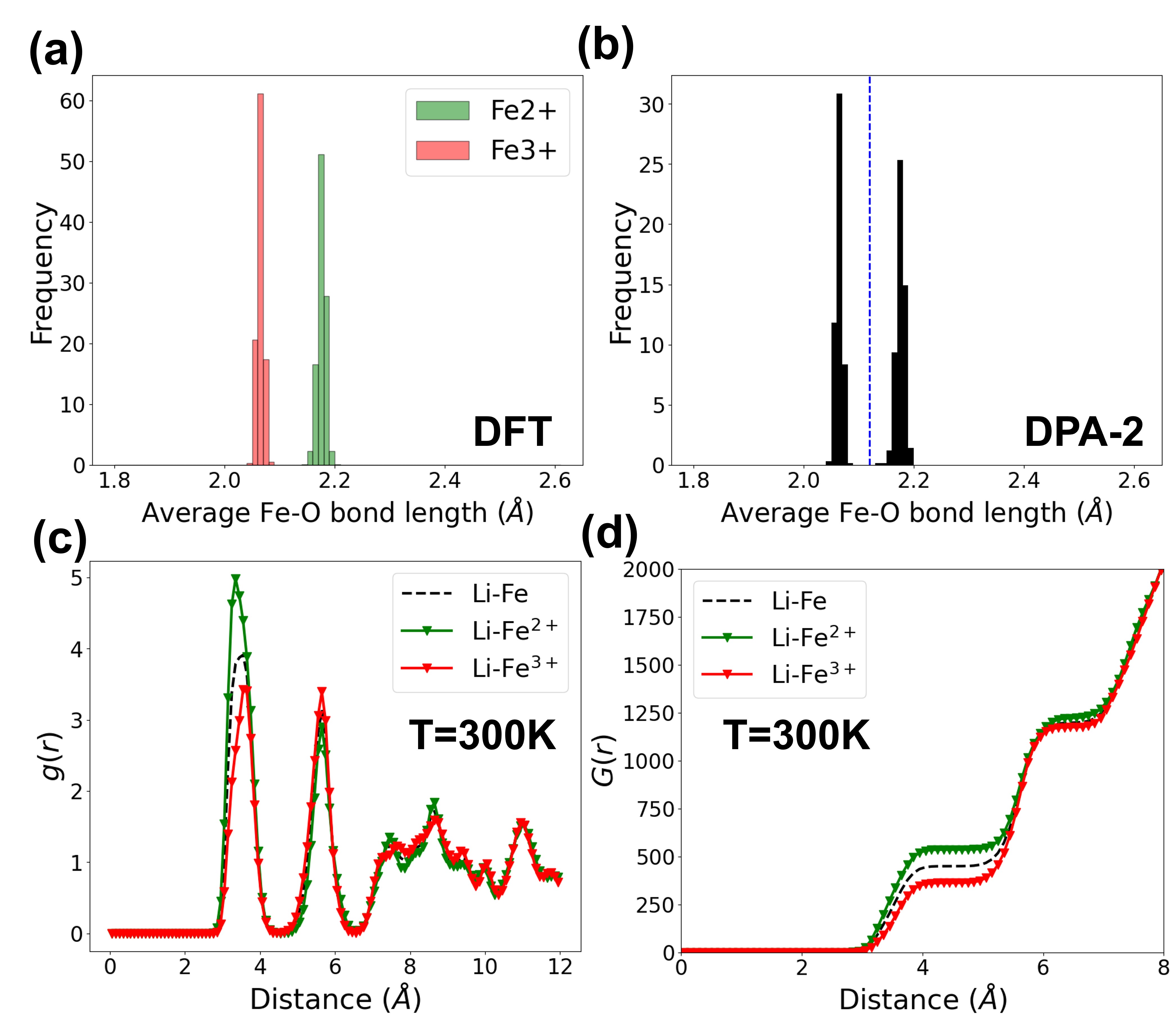}
    \caption{(a) Distribution of Fe ionic radii obtained from DFT-relaxed structures spanning a range of Li, vacancy, and polaron configurations. Fe$^{2+}$ and Fe$^{3+}$ ions, identified via magnetic moments, are shown in green and red, respectively. (b) Distribution of Fe ionic radii after relaxation using the fine-tuned DPA-2 model, starting from configurations 50 steps prior to convergence in the DFT relaxation trajectories. (c) Pair correlation function $g(r)$ and (d) the integral of pair correlation function $G(r)=\int_0^{r} 4\pi r'^2 g(r')dr'$ between Li$^+$ and Fe species at $x=0.50$ and $T=300$ K. Green and red lines represent correlations with Fe$^{2+}$ and Fe$^{3+}$, respectively; the black dashed line shows the average correlation with all Fe ions.}
    \label{fig:fig-s2}
\end{figure}

To capture the energetics associated with small polaron configurations, we performed DFT relaxations on Li$_x$FePO$_4$ structures spanning various compositions $x$ and a range of Li, vacancy, and polaron configurations. Assuming high-spin electronic states, Fe atoms with magnetic moments between 3.6–3.8 $\mu_B$ were identified as Fe$^{2+}$, and those between 4.3–4.5 $\mu_B$ as Fe$^{3+}$. Figure~\ref{fig:fig-s2}(a) shows the distribution of Fe ionic radii, computed as the average Fe–O bond length within each octahedron. The Fe$^{3+}$ population centers around 2.08 $\mathrm{\AA}$, and Fe$^{2+}$ around 2.16 $\mathrm{\AA}$, forming two well-separated Gaussian peaks with minimal overlap. This strong radius-valence correlation enables the use of ionic radius as a reliable indicator for polaron identification.

We then fine-tuned a pretrained DPA-2 model~\cite{zhangDPA2LargeAtomic2024} on the DFT relaxation and AIMD data, and evaluated its ability to recover DFT-relaxed structures from perturbed ionic coordinates. As shown in Figure~\ref{fig:fig-s2}(b), when relaxing structures from intermediate DFT frames (50 steps prior to convergence), the DPA-2 model reproduced the bimodal Fe$^{2+}$/Fe$^{3+}$ radius distribution with high fidelity. Compared to DFT references (matched using pymatgen’s StructureMatcher~\cite{ongPythonMaterialsGenomics2013}), the recovery rate of valence configurations reached 98.4\%. When starting from randomly perturbed structures with 0.05 Å and 0.12 Å atomic shifts, the recovery rates were 100\% and 78.0\%, respectively.

In Figure~\ref{fig:fig-s2}(c) and (d), we further examined spatial correlations between Fe valence states and Li ions at $x=0.50$ and $T=300$ K. Molecular dynamics simulations using the fine-tuned DPA-2 model show that Fe$^{2+}$ exhibits a positive correlation with nearby Li$^{+}$ at 3.5 $\mathrm{\AA}$~distance, while Fe$^{3+}$ shows an anti-correlation, reflecting the expected attractive and repulsive interactions between Li$^{+}$–Fe$^{2+}$ and Li$^{+}$–Fe$^{3+}$ pairs\cite{zhouConfigurationalElectronicEntropy2006, wangDecisiveRoleElectrostatic2023}.

Together, these results demonstrate that the DPA-2 model can accurately reproduce the energetics and spatial features of small polarons in multivalent Li$_x$FePO$_4$ even without explicit charge encoding. This provides a robust foundation for the investigations presented in our work.

\subsection{\label{sec:persistence}Persistence of valence separation at high temperatures}
\begin{figure*}[htbp]
    \centering
    \includegraphics[width=0.9\linewidth]{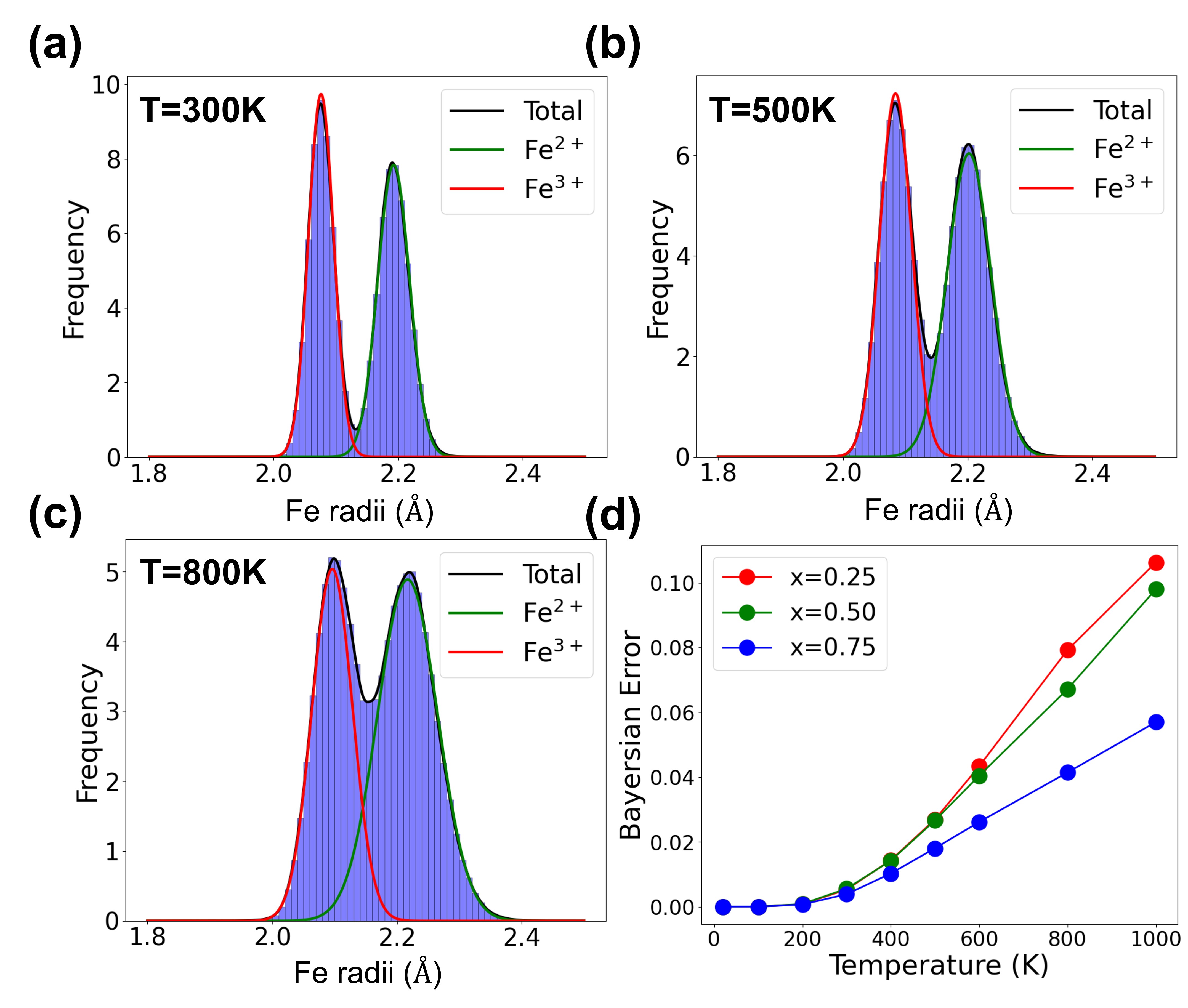}
    \caption{(a–c) Distributions of Fe ionic radii at $x = 0.50$ and $T = 300$, $500$, and $800$ K, respectively. Green and red curves indicate Gaussian fits to the Fe$^{2+}$ and Fe$^{3+}$ subpopulations. (d) Bayesian classification error of Fe valence states as a function of temperature and composition ($x$).}
    \label{fig:fig-s3}
\end{figure*}

To investigate the thermal stability of Fe valence separation, we performed molecular dynamics simulations using the fine-tuned DPA-2 model, starting from 300 K and increasing up to 1000 K to avoid lattice melting. Figures~\ref{fig:fig-s3}(a–c) present the distributions of Fe–O coordination radii at $x = 0.50$ and $T = 300$, 500, and 800 K, respectively. Despite increasing thermal fluctuations, the distributions remain distinctly bimodal up to 800 K, with well-separated peaks corresponding to Fe$^{2+}$ and Fe$^{3+}$ species. As shown in Figure~\ref{fig:fig-s3}(d), we computed the Bayesian classification error (see Section~\ref{sec:post}) using a two-component Gaussian mixture model as a metric of valence-state distinguishability. Although the classification error increases monotonically with temperature, it remains below 0.1 up to 900 K across all compositions examined. These results demonstrate that polaronic Fe$^{2+}$/Fe$^{3+}$ states do not undergo intrinsic thermal collapse, but instead persist dynamically over a wide temperature and compositional range.

\bibliography{refs}